# CW PERFORMANCE OF THE TRIUMF 8 METER LONG RFQ FOR EXOTIC IONS

R. L. Poirier, R. Baartman, P. Bricault, K. Fong, S. Koscielniak, R. Laxdal, A. K. Mitra, L. Root, G. Stanford, D. Pearce, TRIUMF, Vancouver, B. C. Canada

Abstract

The ISAC 35 MHz RFQ is designed to accelerate ions of A/q up to 30 from 2keV/u to 150keV/u in cw mode. The RFQ structure is 8 meters long and the vane-shaped rods are supported by 19 rings spaced 40 cm apart. An unusual feature of the design is the constant synchronous phase of -25°; the buncher and shaper sections are eliminated in favor of an external multi-harmonic buncher. All 19 rings are installed with quadrature positioning of the four rod electrodes aligned to +/- 0.08 mm. Relative field variation and quadruple asymmetry along the 8 meters of the RFQ was measured to be within +/- 1%. Early operation at peak inter-electrode voltage (75kV) was restricted by the rapid growth of dark currents due to field emission; the nominal operating power of 75 kW increased to 100 kW in a few hours. A program of high power pulsing, followed by cw operation have all but eliminated the problem leading to a successful 150 hour test at full power. Successful beam test results confirm beam dynamics and rf designs.

## 1 INTRODUCTION

The accelerator chain of the ISAC radioactive ion beam facility includes a 35.3 MHz split ring RFQ, operating in cw mode, to accelerate unstable nuclei from 2 keV/u to 150 keV/u. The RFQ structure is 8 meters long and the vane-shaped rods, comprised of 40cm long cells, are supported by 19 rings. Full power tests on a single module [1] and on a three-module assembly [2] enabled us to complete the basic electrical and mechanical design for the RFQ accelerator. An initial 2.8m section [3] of the accelerator (7 of 19 rings) was installed and aligned to allow preliminary rf and beam tests to be carried out. The full complement of 19 rings shown in Fig. 1 has now been tested to full rf power with beam.

## 2 DESIGN CONSIDERATIONS

The design of the RFQ is dominated by three considerations. Firstly, the low charge-to-mass ratio of the ions dictate a low operating frequency to achieve adequate transverse focusing. Secondly, continuous wave (cw) operation is required to preserve beam intensity. Thirdly, the desire to minimize the length of the structure and its cost. No single feature of this RFQ gives it exceptional status, but the combination of novel features and unusual design parameters adopted to address these considerations can be argued to give it "landmark" status.

The relative tuning difficulty of an RFQ scales roughly as the vane-length, $L$, divided by the free-space wavelength $\lambda$; for ISAC this ratio is $\approx 1$ which is typical of RFQs. However, the alignment difficulty scales as $L/r_0$ where $r_0$ is the bore radius; and the structure length $L=8$ m of ISAC makes this aspect unusually challenging. The vane voltage of 75 kV is moderate, and the electric field limitation comes not from consideration of the Kilpatrick factor, but rather from the c.w. requirement and cooling limitations. Both these considerations feed into the challenge and complexity of the mechanical design regarding stiffness, stability and tolerances. Though there are several c.w. proton RFQs and one light-ion RFQ, ISAC is unique in c.w. operation for heavier ions. There is no other RFQ operating in cw mode in this frequency range with a charge to mass ratio of 1/30.

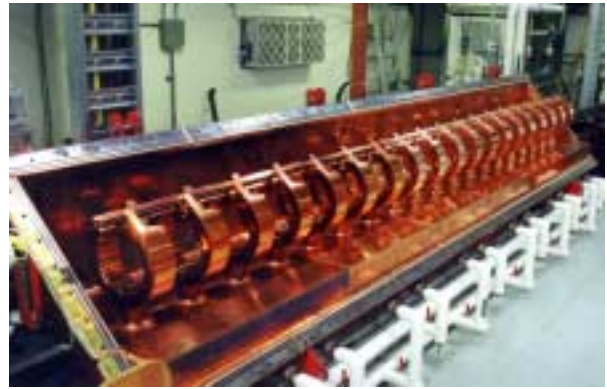

Figure 1. Full compliment of 19 rings installed and aligned

2.1 Beam Dynamics [4]

Though it is now well accepted that a design strategy different from that for high current proton linacs, be used for low current, light and heavy ion RFQs, this was not so at the design time six years ago. The Kilpatrick factor at 1.15 is rather modest. However, because of the power vs. cooling requirement one cannot increase the acceleration rate by merely raising the voltage. In order to reduce the structure length, the buncher and shaper sections were completely eliminated in favour of a discrete four-harmonic saw-tooth pre-buncher located 5m upstream. This has also the benefits of reduced longitudinal emittance at the RFQ exit and of allowing experimentalists to do "time of flight" work with an 86 ns time structure. These gains are made at the expense of a slightly lower beam capture of 80%. Acceleration starts immediately after the radial matching section (RMS) and the vane modulation index ($m$) ramps quickly from 1.124 to 2.6, while the bore shrinks from 0.71 to 0.37 cm in the remaining booster and accelerator sections. A conventional LANL-type design, as for protons, would have resulted in a 12 meter long linac.

To maintain reasonable acceptance, the vane design has characteristic radius to pole tip $r_0 = 0.741$ cm. Though one could single out $m$ and $\phi_s$ as unusual, it is the combination of parameters, chosen so as to hasten acceleration (particularly in the early cells), which is remarkable. The focusing parameter B=3.5, which is "low to typical" of ion-RFQs, is carefully balanced against a comparatively large peak RF-defocusing parameter $\Delta = -0.0408$. For RFQs in general, $\phi_s$ rarely exceeds $-30^o$ and $m$ is rarely above 2 while in ISAC the synchronous phase $\phi_s = -25^o$ is large and constant which maximizes the acceleration and $m=2.6$ which is a record for operational RFQs. Here we adopt the definition $\phi_s = -90^o/0^o$ gives min/max acceleration.

There are several other features of the beam dynamics and vane shapes, which at the time of design were considered quite novel. An exit taper was substituted by a much shorter transition cell, and a transition cell was introduced between the RMS ($m=0$) and the booster. Both entrance and exit region fields of the RFQ vanes were modeled with an electrostatic solver. To minimize machining costs, vanes with constant transverse radius of curvature $\rho = r_0$ were adopted; this leads to significant departure (up to 35% for ISAC) from the two term potential either where $ka \sim$ unity and/or where $m$ is large. Here $k=2\pi/(\beta\lambda)$ and $a$ is the local minimum bore radius. The cell parameters $a,m$ were systematically corrected to compensate for this effect.

The RFQ is also unusual in that the vanes are rotated 45 degrees from the usual horizontal/vertical orientation. Matching into and out of the RFQ therefore requires a round beam. The matching into the RFQ is achieved by four electrostatic quadrupoles. They are the same design as the other quadrupoles in the beam transport line except for the last one. In order to retain an acceptance of greater than 100 $\pi$ mm-mrad through the matching region this quadrupole is very small (1 inch long by 1 inch inside diameter).

## 2.2 RF

From a structural point of view, the low frequency of the RFQ dictates that a semi-lumped resonant structure be used to generate the required rf voltage between the electrodes. Various RFQ models were built [5] and the structure proposed for the ISAC-I accelerator, is a variant of the 4-rod structure developed at the University of Frankfurt [6]. The unique design feature of this 4-rod structure is the single split ring rather than a separate ring for each pair of electrodes resulting in only 8% of the power being dissipated in the tank walls, negating the need for water cooling the tank walls. The choice of the split ring design along with the choice of making $\rho = r_0$ has negated the requirement for individual stem tuners. This type of structure was chosen for its relatively high specific shunt impedance, its mechanical stability and the elimination of an unwanted even-type transmission line mode in favour of the desired odd-mode. The parasitic even-mode was identified as cause of a serious loss of transmission in the HIS RFQ [7] at GSI.

## 2.3 Mechanical

The mechanical design of the RFQ [8] has two major unique features: (1) The vacuum tank is square in cross-section and split diagonally by an "O" ring flange into two parts, the tank base and the tank lid. Each part was plated separately using the triangular enclosure as the container for the cyanide bath solution. In this configuration, the copper plating is easier and it gives full unobstructed access to the RFQ modules for ease of installation and alignment. (2) The basic design of the RFQ structure is different from other RFQ structures in that the maximum rf current carrying surface (rf skin) has been de-coupled from the mechanical support structure (strong-back). The rf skin encloses the strong-back but is attached only where it meets the electrode supports, thus allowing for thermal expansion of the skin without loading the structure and causing electrode misalignment.

# 3 ALIGNMENT

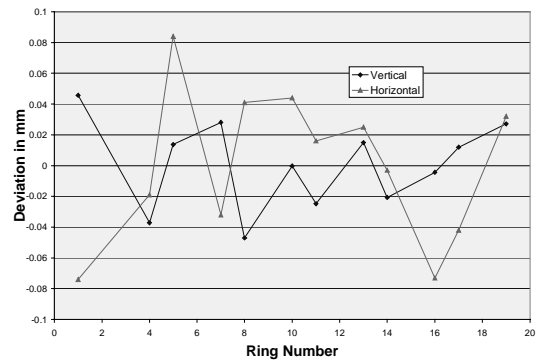

Figure 2. Mechanical alignment of RFQ rings

The alignment goal was to achieve quadrature positioning of the rod electrodes to within 0.08 mm. Beam dynamics calculations indicate that at this level, emittance growth is smaller than 1%. The alignment philosophy [8] was based on manufacturing 19 identical rings and mounting them on precision ground platens, which are accurately aligned in the vacuum tank prior to ring installation. Each platen is special steel 63.5 mm thick with an offset axial rail bolted and doweled in place. The platen and rail are accurately ground in one set-up, thus providing an accurate datum for mounting and locating the ring bases. Each platen has 5 adjustable mounting points, 3 vertical and 2 lateral, and special alignment targets. The platens are adjusted and aligned in the tank using the theodolite intersection method. Once the platens are aligned they are locked in position and ready for installation of the rings. The alignment of the ring assemblies on the platens was accomplished by the same method. Because of the manufacturing procedures and alignment philosophy adopted, when the electrodes were installed on their mounting surfaces they were aligned by definition, assuming that the fabrication tolerances were met.

The three dimensional theodolite technique involves locating two theodolites within a known grid then measuring the angles to monument targets to compute their coordinates. The theodolites and grid lie along one side of the RFQ, and so the horizontals off axis measurements are less accurate because they are close to the theodolite sight line. The alignment results are shown in Fig. 2.

## 4 SIGNAL LEVEL TESTS

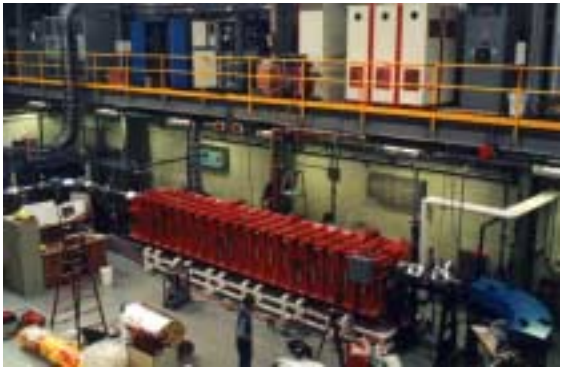

Figure 3. RFQ tank ready for signal level tests

### 4.1 Frequency, Q and Impedance Measurements

Following the mechanical alignment the lid was installed as shown in Fig. 3 ready for signal level tests. The frequency and Q are measured with a network analyzer and the shunt impedance is derived from two independent R/Q measurements; ΔC method and input admittance method [9]. Results are compared to MAFIA calculations in Table 1.

Table 1. Comparison of measured values with calculated MAFIA values.

| Parameter | MAFIA | Measured |
|---|---|---|
| Frequency (MHz) | 34.7 | 35.7 |
| Q | 14816 | 8400 |
| $R_{shunt}$ (k-ohms) | 61.9 | 36.75 |
| $R_{shunt}$ (k-ohms-m) | 470.4 | 279.3 |
| R/Q | 4.18 | 4.375 |

### 4.2 Bead-pull measurements

Since the electrodes have no shoulder upon which to rest the dielectric bead on when measuring the lower gap, the sagging of the bead was overcome by fabricating a bead carriage from teflon that traveled down the center bore of the RFQ and was guided by the straight edges of the electrodes. Nine bead pull runs were made for each set of measurements; carriage only, four separate runs with the dielectric bead in each of the four quadrants and four separate runs by rotating the carriage and the bead together. The carriage only run was used as the average perturbation reference, the run by rotating the carriage with the bead was used to correct any asymmetry in the carriage and the four separate runs were corrected accordingly.

Both the average peak field variation and the quadrupole field asymmetry were deduced from the measurements and are shown in Fig 4. The results are within the target of +/- 1% field strength variation.

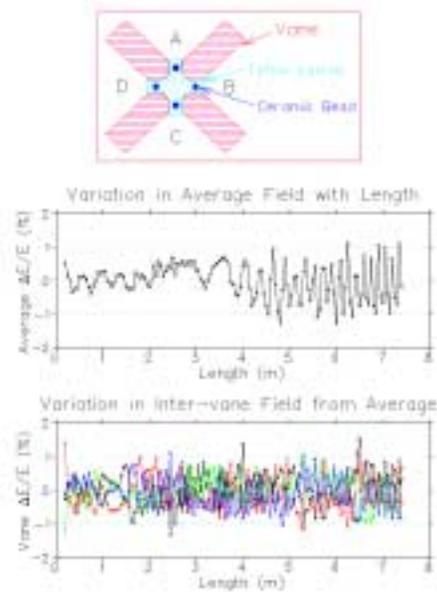

Figure 4. Bead pull measurement results

### 4.3 Transmission line mode measurement.

The decision to install the rf power amplifiers for the ISAC accelerator RF systems several wavelengths away from the RF structures, was based on our experience with matching high Q loads to power sources via a long length of transmission line on other RF systems at TRIUMF [10]. In order to minimize the possibility of a parasitic oscillation at a transmission line mode, it is best for the transmission line electrical length to be a multiple of $\lambda/2$. The parameters of the amplifier tuned circuit and the resonant cavity coupled at each end of the transmission line have an effect on the equivalent electrical length of the transmission line. For the RFQ system, the transmission line resonances were measured to be 2.45 MHz apart. The electrical length of the line was adjusted via a trombone to $n\lambda/2$ indicated by the cavity resonance $f_o$ being centrally located between the two resonances making the difference from $f_o$ to the first transmission line resonance 1.225 MHz.

This is a necessary adjustment for the stable operation of the rf system.

## 5 FULL POWER TESTS

In preparation for full power tests the RFQ tank was baked out for three days at 60° C by uniformly powering eighty-four 500W heaters on the tank walls, covered with a glass fiber blanket to contain the heat (Fig 5). At the same time 60 degree water was circulated through the structure cooling system. A base pressure of 1.4 *10-7 torr was achieved, which increased to 4.0*10-7 torr with full RF power applied.

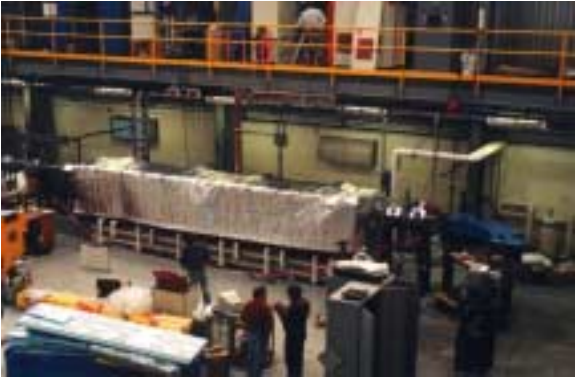

Figure 5. RFQ system ready for bake-out.

Careful cleaning procedures and high power pulsing drastically reduced the growth rate of dark currents associated with field emission. The pulses were 128 us long at a rate of 500 Hz at a peak amplitude of ~100 kV peak. The gradual reduction of dark currents is indicated in Fig. 6 by the reduction of the slope of the sequential graphs, which are in chronological order from top to bottom. Each graph plotted is for a constant voltage after two hours of high power pulsing.

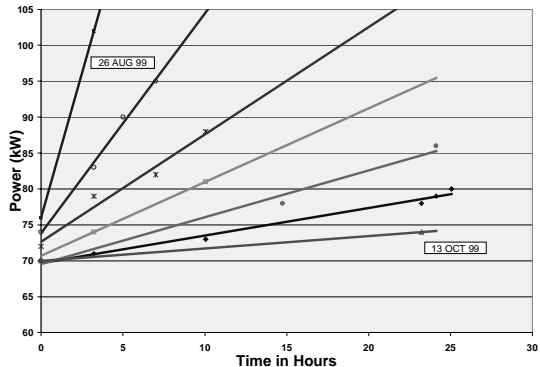

Figure 6. Increase in power level due to dark currents

Initially at the nominal voltage of 74 kV, the dark currents caused an increase of power from 75 kW to 100 kW in 2 hours. Now the power due to dark currents increases by only 5 kW and then levels off in 2 days. A successful 150 hour test at full power was achieved. With no dark currents present, the power requirement to reach an interelectrode voltage of 75 kV is 75 kW. The amplifier is capable of 150 kW for peak power pulsing.

## 6 BEAM TESTS

In 1998 an interim beam test was completed with the first 7 ring section (2.8m) accelerating beams to 55keV/u. In 1999 the final 12 rings were added. Beam commissioning of the complete 19 rings was finally completed this year. The RFQ was operated in cw mode for all beam tests. Beams of $^{4}He^{1+}$, $^{14}N^{1+}$, $^{20}Ne^{1+}$ and $^{14}N_2^{1+}$ all have been accelerated to test the RFQ at various power levels.

In the initial 7-ring test a dedicated test facility was placed downstream of the RFQ. For the final 19-ring configuration the test station was placed downstream of the MEBT and the first DTL section. The test facility includes a transverse emittance scanner, and a 90° bending magnet and Fast Faraday cup for longitudinal emittance estimations.

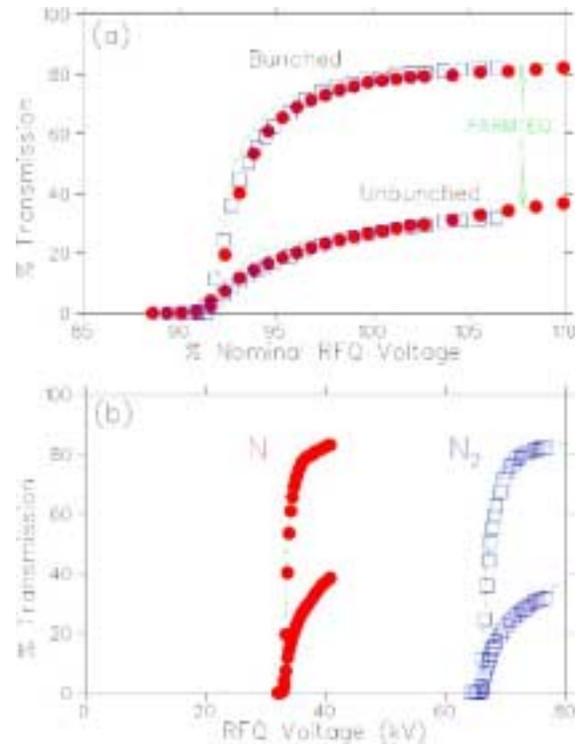

Figure 7. (a) RFQ beam test results showing capture efficiency for beams of $N^+$ as a function of relative vane voltage. The beam capture for both bunched and unbunched initial beams are recorded (squares) and are compared with PARMTEQ calculations (dashed lines). In (b) the results for both $N^+$ and $N^+_2$ are plotted with respect to absolute vane voltage.

Beam capture has been measured as a function of RFQ vane voltage for each ion and for both unbunched and bunched input beams. The MEBT quadrupoles were used as a velocity filter to remove the unaccelerated beam. The results for atomic and molecular Nitrogen are given in Fig. 7 along with predicted efficiencies based on PARMTEQ calculations. The RFQ capture efficiency at the nominal voltage is 80% in the bunched case (three harmonics) and 25% for the unbunched case in reasonable agreement with predictions. A separate measure of the timing pulse train obtained from scattering the beam in a gold foil shows that 5% of the accelerated beam is distributed in the two 35.4 MHz side-bands. This means that after chopping the overall capture efficiency in the 11.8 MHz bunches will be 75%. This will eventually be increased to 80% by adding a fourth harmonic to the pre-buncher.

By varying the MEBT rebuncher while measuring the product of energy spread and time width at an energy or time focus gives an estimate of the longitudinal emittance. The results are shown in Fig. 8 for a $^4He^{1+}$ beam and give an emittance of 0.5 $\pi$ keV/u-ns in agreement with simulations. The measured energy of 153keV/u also is in agreement with design.

Transverse emittances were measured before and after the RFQ. The results show that, when the matching is optimized, the emittance growth in both planes is consistent with zero for the 7-ring configuration for an initial beam of 15π-mm-mrad. In the 19-ring test the emittance scanner was moved after the 90° bend in MEBT. In this case the emittance growth was non-zero but less than a factor of two. It has not been determined what part of the emittance growth is in the RFQ and what part is contributed by the optics.

The transverse and longitudinal acceptances were explored with a so-called 'pencil beam' defined by two circular apertures of 2mm each separated by 0.7m placed in the RFQ injection line. One steering plate was available downstream of the collimators to steer the 'pencil beam' around the RFQ aperture. In the case of the longitudinal acceptance the energy and phase of the incident beam was varied while recording the beam transmission.

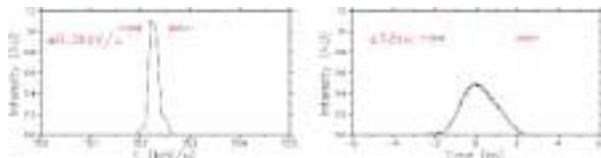

Figure 8. Energy spectrum and corresponding pulse width for an accelerated beam of $^4He^{1+}$.

Based on the steering/transmission data the transverse acceptance was estimated to be ≤ 140 $\pi$ mm-mrad. The longitudinal acceptance was measured for both a centered and an off-centered beam (Ac=2.7 mm) at the nominal RFQ voltage using the pencil beam. The energy and phase settings where the acceptance dropped to 50% of the peak value were used to define the longitudinal acceptance contour. The acceptance of the centered beam was estimated to be 180 $\pi$ %-deg at 35MHz or 0.3π keV/u-ns. The acceptance opens up for off-centered beams with values of 400 $\pi$ %-deg at 35 MHz or 0.7 $\pi$ keV/u-ns. The expected longitudinal acceptance based on PARMTEQ simulations is 0.5π keV/u-ns.

In general the beam test results demonstrate a strong confirmation of both the beam dynamics design and the engineering concept and realization.

## 7 ACKNOWLEDGMENT

We would like to thank Gerardo Dutto, and Paul Schmor, for their valuable technical and managerial discussions.
We are especially grateful to Roland Roper (machine shop) who took on the responsibility of the fabrication and manufacturing details of rings, jigs and fixtures. A special thanks to Bhalwinder Waraich for the mechanical assembly and installation of the rings, and to Peter Harmer for the organization and integration of the RFQ with all the ancillary systems.